\documentclass{article}
\usepackage{arxiv}
\usepackage{authblk}
\usepackage{graphicx}%
\usepackage{multirow}%
\usepackage{amsmath,amssymb,amsfonts}%
\usepackage{amsthm}%
\usepackage{mathrsfs}%
\usepackage[title]{appendix}%
\usepackage{xcolor}%
\usepackage{textcomp}%
\usepackage{manyfoot}%
\usepackage{booktabs}%
\usepackage{algorithm}%
\usepackage{algorithmicx}%
\usepackage{algpseudocode}%
\usepackage{listings}%
\usepackage{csquotes}
\usepackage{xspace}
\usepackage{lineno}
\usepackage{hyperref}
\usepackage{authblk} 
\usepackage[square,numbers,sort&compress]{natbib}

\newcommand{\MPCK}{$\text{MPCK}^\text{+}$}
\newcommand{\EURECA}{EUREC$^4\!\mathrm{A}$\xspace}

\usepackage[singlelinecheck=false]{caption}

\title{Highly Localised Droplet Clustering in Shallow Cumulus Clouds}

\author[1,2]{Birte Thiede}
\author[3,4]{Michael L. Larsen\thanks{\href{mailto:LarsenML@cofc.edu}{LarsenML@cofc.edu}}}
\author[1]{Freja Nordsiek}
\author[1]{Oliver Schlenczek}
\author[1,2,5]{Eberhard Bodenschatz\thanks{\href{mailto:eberhard.bodenschatz@ds.mpg.de}{eberhard.bodenschatz@ds.mpg.de}}}
\author[1]{Gholamhossein Bagheri\thanks{\href{mailto:gholamhossein.bagheri@ds.mpg.de}{gholamhossein.bagheri@ds.mpg.de}}}

\affil[1]{\footnotesize Laboratory for Fluid Physics, Pattern Formation and Biocomplexity, Max Planck Institute for Dynamics and Self-Organisation, Am Fassberg 17, G\"ottingen, D-37077, Germany}
\affil[2]{\footnotesize Institute for the Dynamics of Complex Systems, University of G\"ottingen, Friedrich-Hund-Platz 1, G\"ottingen, D-37077, Germany}
\affil[3]{\footnotesize Department of Physics and Astronomy, College of Charleston, Charleston, SC, 29424, USA}
\affil[4]{\footnotesize Department of Physics, Michigan Technological University, Houghton, MI, 49931, USA}
\affil[5]{\footnotesize Physics Department, Cornell University, Ithaca, NY, 14853, USA}

\begin{document}
\maketitle
\begin{abstract}
The growth, lifetime, number density, and size of water droplets in warm atmospheric clouds determine the evolution, lifetime and light transmission properties of those clouds. These small-scale cloud properties, in addition to precipitation initiation, have strong implications for the Earth's energy budget since warm clouds cover large geographic areas. Spatio-temporal correlations on the millimetre scale and smaller may or may not affect these properties of clouds. To date, the pioneering measurements of such correlations in marine stratocumulus clouds have relied on averaging over holographically reconstructed volumes spanning at least ten kilometres. These have revealed weak but widespread spatial clustering of cloud droplets. Here we present results of strong localised clustering on scales of half a metre or less from holographic measurements collected with the Max Planck CloudKite in shallow cumulus clouds in the mid-Atlantic trade wind region near Barbados, with a spatial separation of only 12~cm between measurement volumes. This observation challenges the foundations of our understanding of cloud microphysics at the droplet scale, with implications for cloud modelling in weather and climate prediction.
\end{abstract}


\begin{figure}[ht]
    \centering
    \includegraphics[width = 1.0\textwidth]{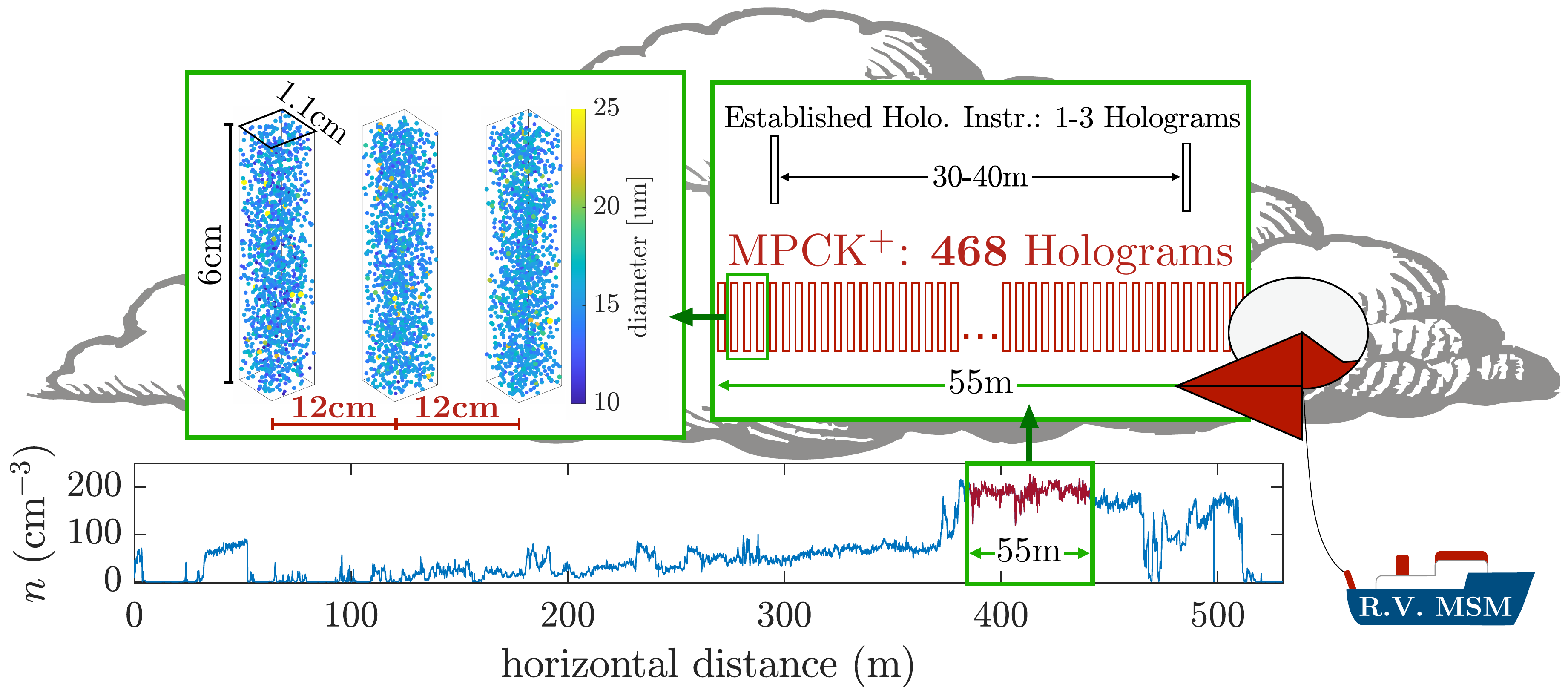}
\caption{Direct cloud droplet measurements with the Max-Planck-CloudKite\textsuperscript{+} instrument box during the \EURECA campaign \cite{stevens21} launched from the Research Vessel Maria S. Merian (R. V. MSM, cruise number 89). Examples of three every 12~cm horizontally spaced holographically recorded droplets (positions and sizes)  within a 1.1~cm $\times$ 1.1~cm $\times$ 6~cm volume are shown on the left top. The instrument box was operated below the CloudKite tethered balloon with a holographic imaging system running at 75~Hz at a TAS of 9~m/s (right). The flight height was $(847 \pm 20)$~m above sea level. The cloud base was at approximated 750~m and the cloud top at 1200-1800~m in the investigated region as observed by radar soundings from  the research vessel \cite{Acquistapace2022}. The droplet number concentration over $500 m$ of measurements is shown on the bottom. For the marked $55$ m of measurement the mean number droplet concentration as measured from the 468 holograms was around $200~\text{cm}^{-3}$ and measurement height only varied by $\pm$5~m. Top right, existing airplane-mounted holographic measurements \cite[e.g.][]{beals15, larsen18c, allwayin24} operating at $\sim$3~Hz and a TAS of $\sim$100~m/s would have one or two measurement in the same $55$~m of cloud.}
    \label{fig:overview}
\end{figure}

Clouds play a key role in determining the global energy balance and regulating the climate system; understanding their formation and interactions with atmospheric circulations is essential for accurate climate simulations \cite{bony2015}.
Although typically studied and modelled at macroscopic scales, the underlying processes that affect cloud formation, evolution, lifetime, and overall radiative forcing are influenced by microphysics driven by interactions at scales of individual droplets \cite{kostinski01a,shaw03}, which are strongly intertwined with the highly turbulent atmospheric flow\cite{bodenschatz2010} and entrainment \cite{beals15,bodenschatz15}. The local environment of a cloud droplet (including the presence or absence of other nearby droplets and their corresponding influence on temperature and vapour fields) is expected to have a significant impact on microphysical processes \cite{shaw03,grabowski2013,pumir16}. 
Turbulence not only leads to complex mixing and transport of aerosols, temperature, and vapour, but may also have a yet to be understood effect on the inertia-driven clustering and collision-coalescence of cloud droplets.  Our limited understanding of microphysical processes is exemplified by our inability to fully explain how rain initiates in warm clouds, which is an unsolved mystery that has puzzled scientists for many decades \cite{langmuir1948, saffman1956,telford80,beard93,yau96,feingold99,falkovich02,vaillancourt02,shaw03,blyth03,kostinski05, toschi09, morrison20, li22}.

Efforts to characterise inertial particle clustering in clouds have been ongoing for at least 30 years, often using \emph{in situ} measurements of cloud particle spatial positions \cite{baker92,baumgardner93,uhlig98,kostinski00,chaumat01,kostinski01b}. Typically, observations were made with aircraft-mounted probes that detected particle sizes and arrival times within a narrow optical field-of-view, enabling the reconstruction of one-dimensional spatial positions based on airspeed. Inferring three-dimensional clustering from such quasi one-dimensional data required assumptions about statistical isotropy and stationarity \cite{holtzer02,larsen14b}.
Digital in-line  holography reconstructs true volumetric droplet spatial positions and sizes. Early efforts date back decades \cite{conway82,kozikowska84,brown89,uhlig98}, but only recent advances in hardware and software enable high-resolution reconstructions.
One prominent instrument, the Holographic Detector for Clouds (HOLODEC), reconstructs up to $\sim 10$ cm$^3$ cloud volumes and provides the most spatially resolved \emph{in situ} picture of three-dimensional cloud microstructure to date \cite{fugal04,fugal09,spuler11,beals15,glienke17,glienke20}. Nevertheless, characterising scale-dependent clustering remained challenging due to the limited number of droplets per holographic volume. By combining data from hundreds of consecutive holograms taken while flying through clouds, researchers detected weak but statistically significant millimetre-scale clustering averaged over 30+km long stretches of stratocumulus clouds \cite{larsen18c}.

An important and often used tool to quantify statistically droplet clustering in clouds is the three-dimensional radial distribution function (RDF, or $g(r)$, see Eq. S4 in the Supplement), which measures deviations from perfect spatial randomness, where $g(r_\circ) = 1$ indicates randomness at the scale $r_\circ$ and $g(r_\circ) = 1. 3$, for example, means that particles separated by $(r_\circ,r_\circ + \text{d}r)$ are found 30\% more often than expected for a perfectly random system with the same underlying particle density.
The radial distribution function is only statistically meaningful when examining statistically homogeneous and stationary data \cite{shaw02a,larsen05,kostinski06,larsen06,larsen07}. Caution in interpreting the RDFs is needed for non-homogeneous and  non-stationary situations \cite{larsen12}. The analysis of \citet{larsen18c} was based on the assumption that hundreds to thousands of consecutive holographic volumes collected over tens of kilometres of cloud were statistically homogeneous and stationary. This was justified by the relatively constant observed drop number concentration and size distribution throughout the data. As we show below, the assumption of statistical homogeneity and stationarity over such large scales may be physically questionable and may have influenced the interpretation of the results. 

Here we report holographic measurements with the Advanced Max Planck CloudKite instrument (\MPCK) instrument inside shallow cumulus clouds during the \EURECA campaign in the Atlantic trade wind regions near Barbados \cite{stevens21}.  As shown in Fig. \ref{fig:overview}, the high acquisition rate of the \MPCK holography system of 75 Hz, carried on the tethered helikite at a low true air speed (TAS) of $\sim$9 m/s, enabled unprecedented spatio-temporal volumetric measurements of shallow cumulus cloud droplets  with an interhologram distance of only 12 cm. In addition, the particle image velocimetry system of the \MPCK  that observed the same volume of air as the holographic imager allowed us to eliminate possible systematic errors due to the mounting of the instrument under the helikite. Further analysis also confirms that the holographic arms do not influence the results, as detailed in the supplementary information.

\begin{figure}
    \centering
    \includegraphics[width=0.6\columnwidth]{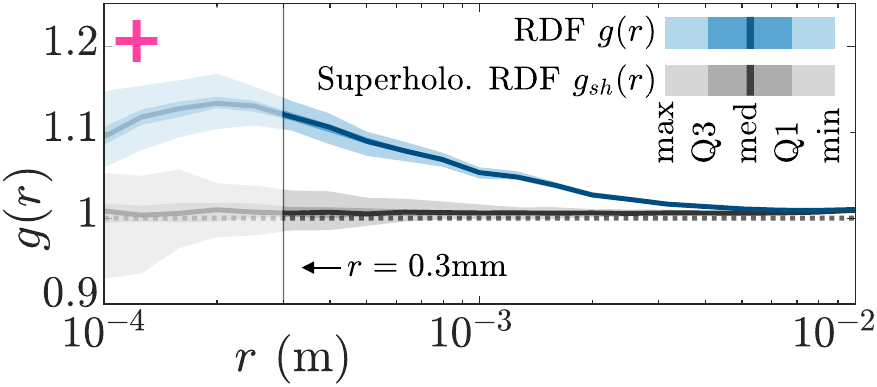}
    \caption{The radial distribution function averaged over the 468 holograms. Deviations from $g(r)=1$ can be due to both real, physical clustering and statistical/instrumental uncertainties; the gray line and shadings quantify some of these uncertainties. The gray bounds are calculated from artificial holograms, where observed hologram cloud droplet number variability is retained and droplet positions are randomly sampled from all droplet positions observed throughout the 55m measurement interval (see supplement for more detail). The blue envelope shows $g(r)$, the RDF from actual observed droplet position data perturbed by the depth position uncertainty $\pm$ 150 $\mu$m. Due to measurement inaccuracies, only $r>0.3$~mm is reliable; smaller distances are shown blurred. The regions where the two envelopes do not overlap (here on scales up to approximately 5 mm) show evidence for statistically significant droplet clustering throughout the interval. This approach, and the qualitative shape of this RDF curve, are largely consistent with the results published for marine stratocumulus clouds \citet{larsen18c}.}
    \label{fig:fullrdf}
\end{figure}

This study analyses the clustering signature in a segment of data collected within a precipitating shallow cumulus cloud. The selected data segment includes 468 consecutive holograms, covering a $\sim$55-metre horizontal cloud region (Fig. \ref{fig:overview}). Following \cite{larsen18c,glienke20}, we focused on a cloud region with relatively stable droplet number concentration, liquid water content, and mass-weighted mean diameter. During this interval, the \MPCK altitude varied by less than 10 m.
For consistency with earlier studies, the analysis was restricted to droplets with a minimum diameter of 10~$\mu$m within a 1.1~cm x 1.1~cm x 6.0~cm core region of the holograms. In this sample volume, a ``superhologram'' heat map showed no visual or statistically obvious deviations from uniform drop detection efficiency (see also supplement). Additional details about the field study, instruments, measurement platforms, hologram post-processing technique and verification, and radial distribution function calculations can be found in the supplement.

For calculating the three-dimensional RDF from the finite-volume holograms, earlier methods for calculating the RDF \cite{larsen18a} were refined, rigorously tested, and validated using simulated datasets (see Supplement for details). The RDF for the entire 55-metre domain is shown in Fig. \ref{fig:fullrdf}, derived from all inter-particle distances within the 7.26~cm\textsuperscript{3} volumes of the 468 holograms collected. As noted, values of $g(r) > 1$ indicate a higher-than-perfectly-random frequency of inter-particle distances at a given scale $r$, i.e. clustering at scale $r$.
The uncertainty in $g(r)$ may be influenced by multiple factors including sample size, measurement volume aspect ratio, droplet number concentration, and spatial scale $r$. Additionally, droplet data from any instrument has biases in detection and position accuracy. Thus, the RDF results require a meticulous interpretation, which is given below.

Figure \ref{fig:fullrdf} shows two main traces: the blue line $g(r)$ calculated from averaging over individual holograms and the gray line $g_{sh}(r)$ from randomly drawing from pooling droplet data from all holograms, i.e. sampling from a \enquote{super hologram}. Ideal, unbiased measurements would yield $g_{sh}(r) = 1.0$ at all scales, but here minor deviations appear due to finite sampling and/or instrumental sensitivity limitations.
To account for uncertainties in reconstructed droplet positions, $g(r)$ and $g_{sh}(r)$ were calculated for 100 resamplings, with positions varied within quantified uncertainties \cite{methodspaper}, e.g. $\pm150$~$\mu$m along the camera axis. Solid lines show the median of these resamplings, and shaded areas span the minimum, maximum, and inter-quartile ranges for the 100-member ensemble. Given positional inaccuracies on the order of $10^{-4}$m, retrieval of RDF statistics for spatial scales less than 0.3~mm were too uncertain to be reliable (further details see Supplement).
\begin{figure}
    \centering
    \includegraphics[width=0.6\columnwidth]{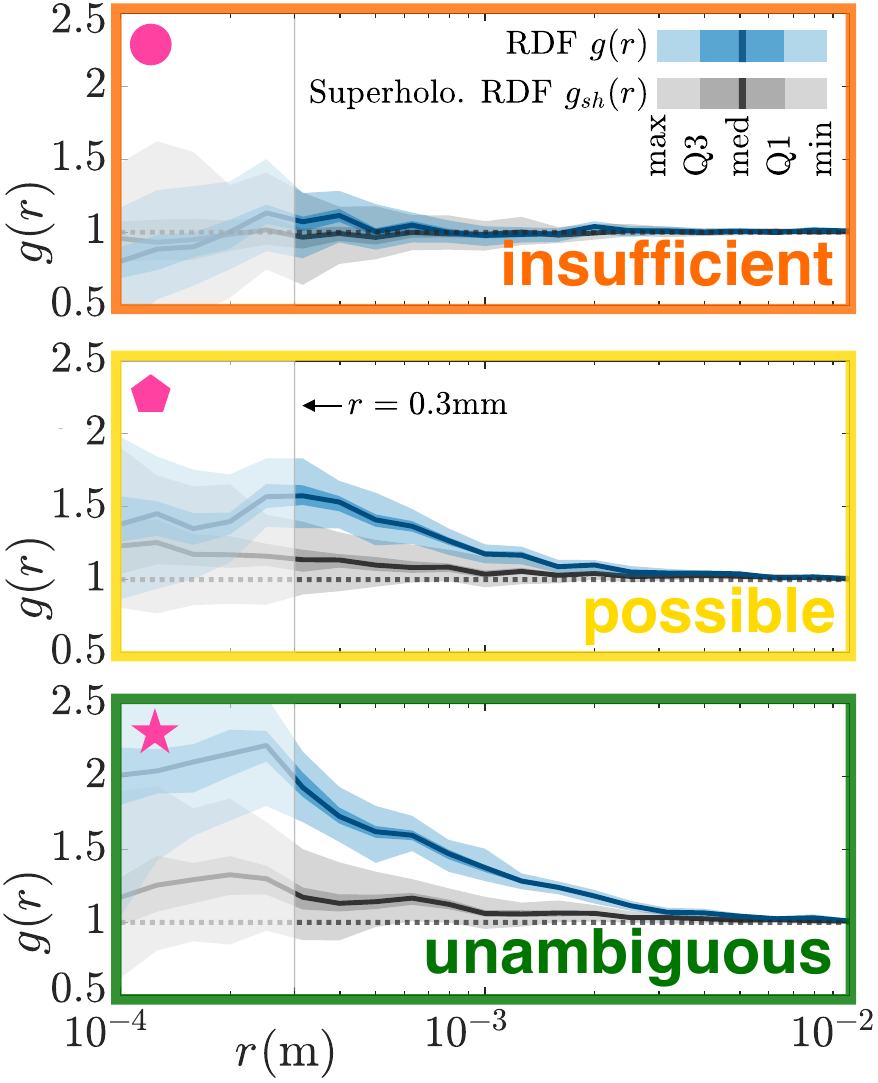}
    \caption{Examples of RDF-envelopes based on 4 consecutive holograms each. As described in figure \ref{fig:fullrdf}, the RDF of measured droplets are shown with a blue envelope whereas the gray envelope serves as a reference revealing artefacts from sampling uncertainties and possible instrumental anomalies. For regions where $g(r)$ and $g_{sh}(r)$ envelopes fully overlap the clustering evidence is classified as \textbf{insufficient}, if separation only occurs for a few $r$ values we classify clustering as \textbf{possible}, if, however,  they separate for at least 4 different $r$ values we classify the evidence for clustering as \textbf{unambiguous}. }
    \label{fig:clustertype}
\end{figure}

Consistent with the argument given in \citet{larsen18c}, we argue there is evidence for droplet clustering if the envelope corresponding to the actual data $g(r)$ lies outside the envelope corresponding to the super hologram $g_{sh}(r)$. The width of the envelopes at each scale help indicate the magnitude of the uncertainty of the radial distribution function. If at $r_\circ$ the RDF-envelope $g(r_\circ) = 1+(\gamma \pm \delta \gamma)$ and the superhologram-RDF-envelope $g_{sh}(r_\circ) = 1+(\Gamma \pm \delta \Gamma)$, then the value of the actual RDF $g(r_\circ)$ for the measured droplets can be estimated to be $g(r_\circ) \sim 1 + (\gamma-\Gamma) \pm (\delta \gamma + \delta \Gamma)$.
The results shown in figure \ref{fig:fullrdf} qualitatively agree with the major findings of \citet{larsen18c}; RDFs show net spatial clustering (the blue envelope lies above the gray), decrease monotonically with increasing spatial scale, and exhibit clustering near the turbulence dissipation range. Quantitatively, however, the clustering observed here is about 5-10 times stronger than reported in \citet{larsen18c}.
 The difference in the magnitude of the clustering signature is surprising at first sight; is this due to different cloud types (marine shallow cumulus vs.\ marine stratocumulus) and/or perhaps due to more intense turbulence? Could it be that the distance of 30~m between measurements in \citet{larsen18c} averaged out the observed clustering? Thanks to the very high spatial sampling of only 12~cm distance between holograms by the \MPCK we can probe the RDF at smaller averaging scales, i.e. subdividing the interval into very fine domains. This is possible here due to the sufficient number of drops per hologram due to the large sample volume and high droplet concentration.   Figure~\ref{fig:clustertype} illustrates this approach using sub-domains of four consecutive holograms taken at locations within the 468-holograms (55~m). These 4-hologram sub-domains average only over $\sim$0.5~m of cloud. Each subplot compares the measured $g(r)$ envelope (blue) with the associated super-hologram $g_{sh}(r)$ envelope (gray). Compared to the full-domain envelopes in Figure~\ref{fig:fullrdf}, these smaller sub-domains show substantially wider envelopes, indicating the higher uncertainty expected from detecting fewer droplets.The width of scales in which the measured RDF is outside of the envelopes of the super-hologram RDF allows us to classify our RDF as \enquote{insufficient}, \enquote{possible}, or \enquote{unambiguous}. Examples of each are shown in Figure~\ref{fig:clustertype}. For instance, although $g(r = 0.3~\text{mm}) \approx 1.1$ in the top panel is comparable to the full-domain value in Figure~\ref{fig:fullrdf} -- where clustering evidence was found -- the increased uncertainty in this 4-hologram subset (demonstrated by the thickness of the associated envelopes) makes it impossible to draw the same conclusion here. Thus, the classification \enquote{insufficient} was chosen instead of \enquote{no clustering}; there may indeed be clustering, but the limited droplet count precludes confident determination. The middle subplot shows $g(r)$ and $g_{sh}(r)$ for another 4-hologram sample, where the envelopes do not overlap in a small region near $r = 0.4\ \text{mm}$, and clustering is  \enquote{possible}. In contrast, the lower subplot of a different 4 hologram sample reveals a clear separation of envelopes over a significant range of scales and clustering evidence is thus \enquote{unambiguous}.

 \begin{figure}[!h]
    \centering
    \includegraphics[width = 1.0\textwidth]{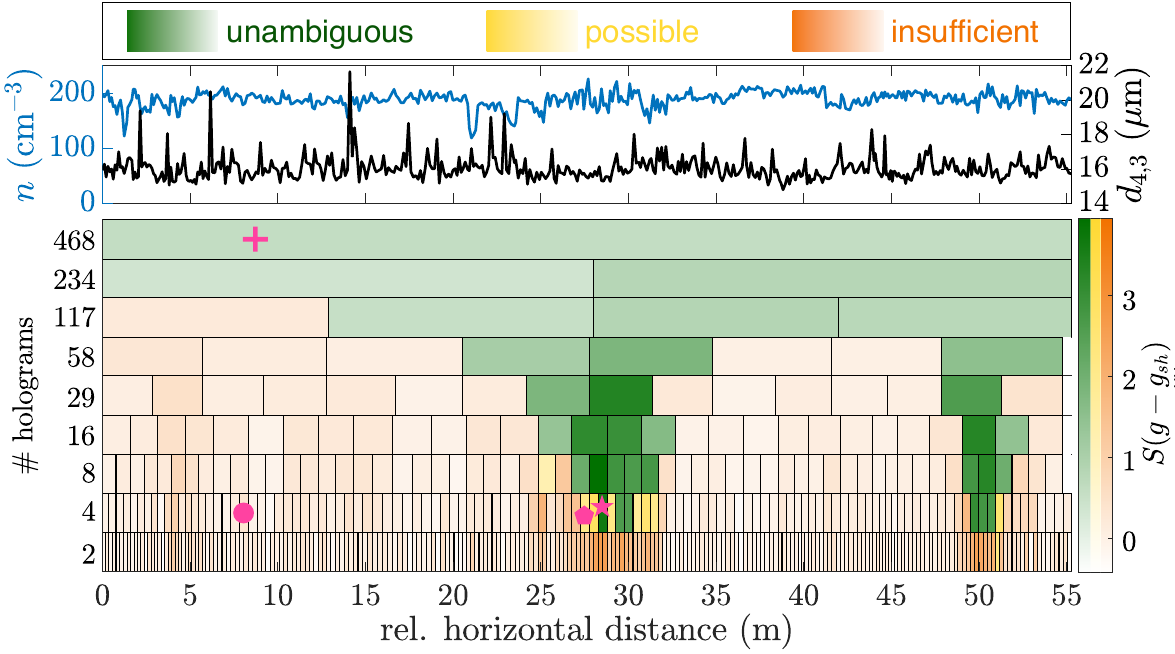}
    \caption{An overview of where clustering evidence can be found within subdomains of the $\sim$55 m data. The top panel shows traces of the droplet number concentration $n$ and the mass-weighted mean diameter $d_{4,3}$. The bottom panel shows which of the three clustering evidence classes each sub-interval is classified into. Green indicates unambiguous evidence of statistically significant droplet clustering. The marked sub-intervals (plus, circle, pentagon, and star) correspond to the shown RDFs in figures \ref{fig:fullrdf} and \ref{fig:clustertype}. The colour opacity show a measure of RDF amplitude, computed by the sum of the difference of $g(r)$ and $g_{sh}(r)$ across scales, i.e. $S(g - g_{sh}) = \sum_{r \in \{0.3\,\mathrm{mm}, \dots, 10\,\mathrm{mm}\}}\bigl[\mathrm{med}(g(r)) - \mathrm{med}(g_{sh}(r))\bigr]$, where \enquote{med} stands for the median. $S(g - g_{sh})$ is a crude measure of the RDF amplitude above the uncertainty level. Note that some of the strongest RDF amplitudes seem to occur even where there is ``insufficient'' evidence for clustering; with 2 holograms, the differences between the superhologram-RDF that make up $g_{sh}(r)$ cannot be reliably distinguished from the measured $g(r)$ values.}
    \label{fig:chart}
\end{figure}
By classifying clustering evidence based on $g(r)$ and $g_{sh}(r)$ across disjoint sub-intervals of the 55-metre transect, our findings reveal that the clustering is far more localised than previously anticipated.
Figure~\ref{fig:chart} highlights (in color) the sub-intervals where clustering is evident.
Figure~\ref{fig:chart} reveals that the modest clustering signal over the entire 55-metre domain is driven by stronger, localised clustering in only a small fraction of the domain. The spatial extent of the clustered region is no more than a few meters in horizontal extent within these shallow cumulus clouds. 
 The measurement uncertainty limits us to averaging of 50-100~cm horizontal extent (4-8 hologram sub-domains) to find a clear clustering signal. At this point it is not clear whether the intermittency would be observable at even smaller scales.
 Clustering evidence at a given scale points to localised sub-scales with stronger clustering signatures (i.e., no “green” shading lies below an “orange” shading in Figure~\ref{fig:chart}). These observations, which align with additional \MPCK\ data from \EURECA (not shown), reinforce that while small-scale clustering may seem pervasive in large domains as in \citet{larsen18c}, it is often strongly localised, with its true magnitude hidden by averaging in prior aircraft measurements.
Notably, these \enquote{clustering hotspots} do not correlate with variations in droplet number concentration, droplet diameter, liquid water content, droplet-size-distribution width, altitude, direction of the wind velocity vector or any other examined variable—indicating the clustering signature is both spatially intermittent and seemingly decoupled from parameters one might first expect to influence or be affected by it.
One can speculate that this observation may be due to the turbulence intensity within the cloud. Future work will survey a more comprehensive ensemble of datasets to characterise these domains in more detail and to further search for any links to other parameters that may correlate, such as the local turbulent energy-dissipation-rate.

In conclusion, the presented high-resolution measurements of cloud droplets in shallow cumulus clouds reveal strikingly strong and intermittent clustering at sub-meter scales, challenging prevailing assumptions of cloud microphysical structure and suggesting that current collisional growth models may require refinement.The drop-drop collision rates expected in these clustering hotspots will be larger than for a perfectly random or weakly clustered cloud, perhaps to an extent that would help solve existing puzzles related to warm rain initiation. Here  further investigation is warranted. The observed clustering intermittency in a precipitating cloud supports the notion that only a few \enquote{lucky} droplets are needed to trigger rain in warm clouds\cite{kostinski05}. Verifying whether these findings genuinely link to rain formation in warm clouds, however,  demands broader observations across diverse warm-cloud regimes, conducted at spatial and temporal resolutions on par with or surpassing those of the \MPCK. The impact of the clustering hotspots reported here on weather and climate prediction remains to be clarified and will be the subject of future research.

\section*{Acknowledgements}
We would like to thank the organisers of \EURECA for making CloudKites' participation in the field campaign possible. We deeply appreciate the dedication of the MPI-DS scientific electronics and machine shop, as well as the captains and crew of R.V. MSM89. Additionally, we extend our gratitude to Johannes G{\"u}ttler and Kashawn Hall for their invaluable assistance in operating the Max Planck CloudKite during the campaign. This research and \EURECA field campaign have been supported by the people and government of Barbados; the Max Planck Society and its supporting members; the German Research Foundation (DFG) and the German Federal Ministry of Education and Research (grant nos. GPF18-1\textunderscore69 and GPF18-2\textunderscore50); the European Research Council (ERC) advanced grant \EURECA (grant agreement no. 694768) under the European Union’s Horizon 2020 research and innovation program (H2020). M.L. was supported by the National Science Foundation under Grant AGS-2001490.

\section*{Conflict of interest}
The authors declare no competing interests. The sponsors had no influence on the study design, data collection and analysis, decision to publish, or preparation of the manuscript.

\section*{Author contribution}
BT and ML contributed equally as the primary authors of this work. BT, ML, EB, and GB conceptualised the work; BT, FN, OS, EB, and GB were responsible for data acquisition and curation; BT, ML, and GB conducted the formal analysis; ML, EB and GB acquired associated funding; EB and GB were the project administrators; BT, ML and GB wrote the original draft; and all authors assisted with review and editing of the final manuscript.

\bibliographystyle{unsrtnat}

\end{document}